\documentclass[12pt,preprint]{aastex}
\usepackage{emulateapj5}
\usepackage{apjfonts}
\usepackage{amssymb, amsmath, natbib}
\usepackage{graphicx}
\usepackage{CJK}

\input psfig.tex

\newdimen\digitwidth      
\setbox1=\hbox{0}       
\digitwidth=\wd1        
\catcode`"=\active      

\def"{\kern\digitwidth}

\def\aj{{AJ}}

\def\apj{{ApJ}}
\def\apjs{{ApJS}}
\def\asec{$^{\prime\prime}$}
\def\deg{$^{\circ}$}

\def\hst{{\it HST}}

\def\lax{{$\mathrel{\hbox{\rlap{\hbox{\lower4pt\hbox{$\sim$}}}\hbox{$<$}}}$}}
\def\gax{{$\mathrel{\hbox{\rlap{\hbox{\lower4pt\hbox{$\sim$}}}\hbox{$>$}}}$}}
\def\simlt{\lower.5ex\hbox{$\; \buildrel < \over \sim \;$}}
\def\simgt{\lower.5ex\hbox{$\; \buildrel > \over \sim \;$}}

\def\mnras{{MNRAS}}
\def\nat{{Nature}}

\def\galfit{{\tt GALFIT}}
\def\ser{{S\'{e}rsic}}

\slugcomment{submitted to The Astrophysical Journal.}
\shorttitle{MINOR MERGER REMNANTS IN NGC~4889}
\shortauthors{Gu et al.}

\begin{document}

\begin{CJK*}{UTF8}{gbsn}

\title{
A Novel Approach to Constrain the Mass Ratio of Minor Mergers in Elliptical 
Galaxies: Application to NGC~4889, the Brightest Cluster Galaxy in Coma
\altaffilmark{1} 
}

\author{Meng Gu (顾梦)\altaffilmark{2,3}, Luis C. Ho\altaffilmark{3}, 
Chien Y. Peng\altaffilmark{4} and Song Huang (黄崧)\altaffilmark{2,3,5}}
\date{}                                          

\altaffiltext{1}{Based on observations made with the NASA/ESA  {\it Hubble
Space Telescope}, obtained from the Data Archive at the Space Telescope
Science Institute, which is operated by the Association of Universities for
Research in Astronomy (AURA), Inc., under NASA contract NAS5-26555.}

\altaffiltext{2}{School of Astronomy and Space Science, Nanjing University,
Nanjing 210093, China}

\altaffiltext{3}{The Observatories of the Carnegie Institution for Science, 
813 Santa Barbara Street, Pasadena, CA 91101, USA}

\altaffiltext{4}{Giant Magellan Telescope Organization, 251 South Lake Avenue,
Suite 300, Pasadena, CA 91101, USA}

\altaffiltext{5}{Key Laboratory of Modern Astronomy and Astrophysics, Nanjing
University, Nanjing 210093, China}

\begin{abstract}

Minor mergers are thought to be important for the build-up and structural 
evolution of massive elliptical galaxies. In this work, we report the discovery 
of a system of four shell features in NGC~4889, one of the brightest members of 
the Coma cluster, using optical images taken with the {\it Hubble Space 
Telescope}\ and the Sloan Digital Sky Survey. The shells are well aligned with 
the major axis of the host and are likely to have been formed by the accretion 
of a small satellite galaxy.  We have performed a detailed two-dimensional 
photometric decomposition of NGC~4889 and of the many overlapping nearby 
galaxies in its vicinity.  This comprehensive model allows us not only to firmly 
detect the low-surface brightness shells, but, crucially, also to accurately 
measure their luminosities and colors.  The shells are bluer than the underlying 
stars at the same radius in the main galaxy.  We make use of the colors of the 
shells and the color-magnitude relation of the Coma cluster to infer the 
luminosity (or mass) of the progenitor galaxy.  The shells in NGC~4889 appear to 
have been produced by the minor merger of a moderate-luminosity 
($M_I \approx -18.7$ mag) disk (S0 or spiral) galaxy with a luminosity (mass) 
ratio of $\sim$90:1 with respect to the primary galaxy.  The novel methodology 
presented in this work can be exploited to decode the fossil record imprinted in 
the photometric substructure of other nearby early-type galaxies.
\end{abstract}


\maketitle

\section{Introduction}

The evolution of massive early-type galaxies has always been a major concern in 
galaxy formation theory.  This topic is becoming increasingly interesting due to 
the recent discovery of quiescent, compact early-type galaxies at high redshifts 
(Cimatti et al. 2004; Daddi et al. 2005; Trujillo et al. 2006).  The population
of ``red nuggets,'' which represent a unique phase of massive galaxy evolution
(van Dokkum et al. 2010), is characterized by much smaller effective radius
(Trujillo et al. 2006, 2007; Damjanov et al. 2011), higher central density 
(Szomoru et al. 2012), and possibly higher stellar velocity dispersion 
(Cappellari et al. 2009; Onodera et al. 2012) compared with nearby early-type 
galaxies at similar stellar mass. A significant fraction of these objects have 
disk-like morphologies (van der Wel et al. 2008).  

Judging from their number density and stellar mass, red nuggets are considered 
probable progenitors of local massive early-type galaxies.  To accomplish this 
transformation, the high-$z$ compact objects on average need to double in 
stellar mass while increasing their effective radius by a factor of $3-5$ since 
$z\approx 2.0$ (van Dokkum et al. 2010). To match the radial profiles of local 
massive elliptical galaxies, an extended stellar envelope needs to be gradually 
accumulated around the compact cores (Bezanson et al. 2009; Szomoru et al. 
2012), whose high density implies a rapid phase of early, dissipative collapse.  
Moreover, most of the late-time growth needs to be accomplished without much 
significant gas dissipation to satisfy the observational constraint that 
early-type galaxies at $z$ \lax\ 2 experience little observable star formation 
or black hole accretion (Bezanson et al. 2009; van Dokkum et al. 2010).  This 
two-phase evolutionary picture strongly challenges conventional models for the 
formation of elliptical galaxies, both monolithic collapse (Larson 1975) and 
binary merger (Toomre \& Toomre 1972) scenarios.  Instead, dissipationless
processes such as ``dry'' mergers play a more central role.  Considering the 
rarity of major (i.e. nearly equal-mass) dry mergers, late-time accretion of 
smaller stellar systems without significant cold gas content---minor dry 
mergers---appears to be the most promising pathway for achieving the inferred 
structural evolution (Naab et al. 2009).  A series of simulations (Oser et al. 
2010, 2012; Lackner et al. 2012) describes the two-phase formation scenario for 
massive early-type galaxies.  According to this scenario, massive galaxies 
experience strong dissipational processes such as cold accretion (Dekel et al. 
2009) and gas-rich (``wet'') mergers at high redshifts, resulting in early, 
rapid, concentrated mass growth. At lower redshifts, accretion of small systems 
dominates the evolution by building up the outer low-density envelopes around 
the initially compact cores (Naab et al. 2006, 2009; Oser et al. 2010). This 
picture has gained support from a variety of observations of nearby massive
ellipticals (Coccato et al. 2010; Greene et al. 2012; Huang et al. 2013).
 
Located at the extreme high-mass end of the stellar mass function and the 
densest environments in the Universe, brightest cluster galaxies (BCGs) 
have been considered a special class of early-type galaxies that can be traced 
and compared across different redshifts (Sandage 1976; Postman \& Lauer 1995).
Motivated by the unique properties of BCGs, early ideas for their formation 
explored the role of X-ray-driven cooling flows (Silk 1976; Fabian 1994) and 
cannibalism (Ostriker \& Tremaine 1975). The cooling flow picture emphasizes 
the im-
\vskip 0.1cm  
\begin{figure*}[tb]
\centerline{\psfig{file=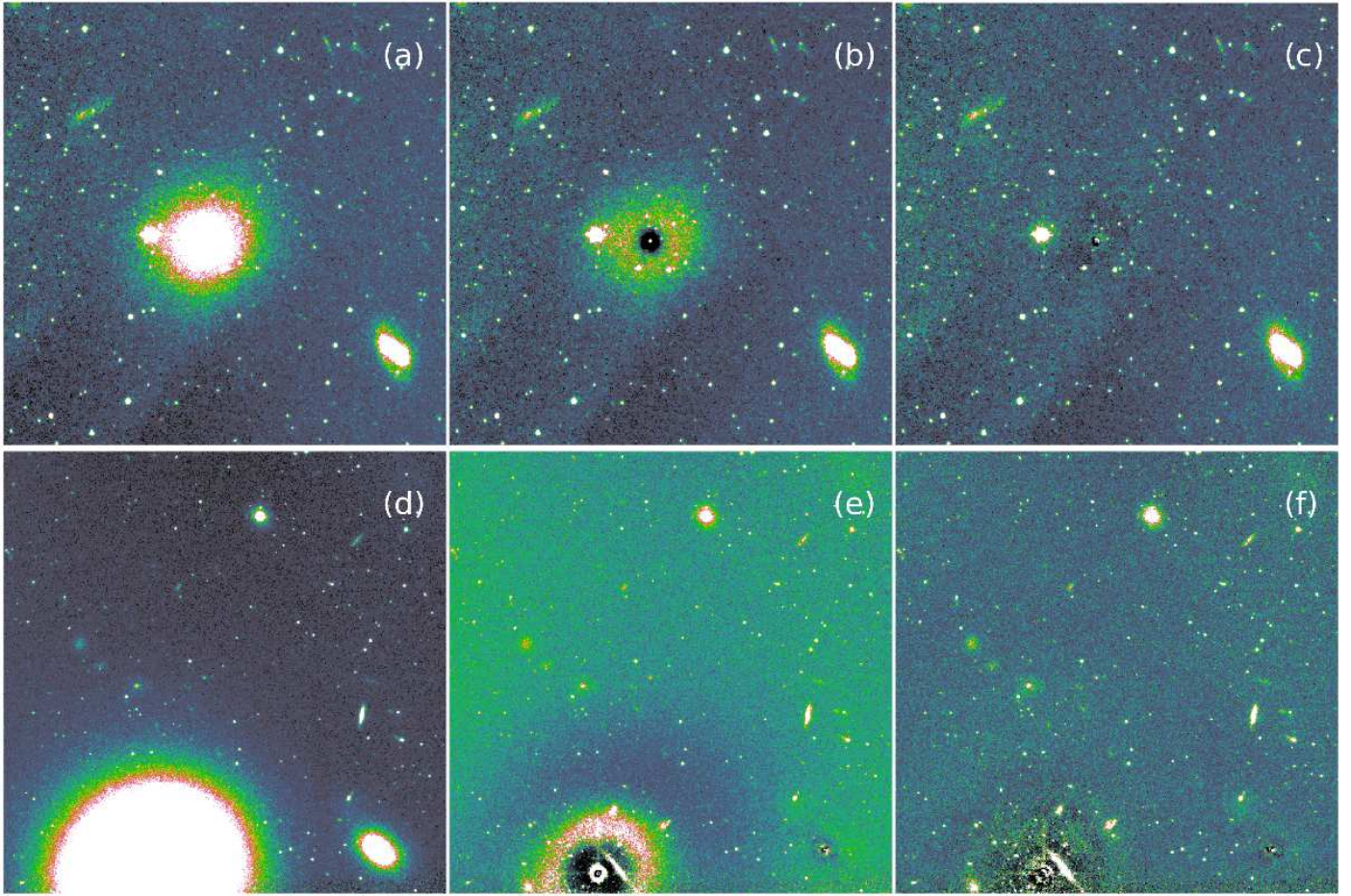,width=18.5cm}}
\figcaption[fig1.eps]{
Examples of multiple-component fitting applied to sections of the 
{\it HST}/ACS F814W image that contain galaxies near NGC~4889.  The top row 
shows a relatively simple galaxy (a), a single-component model (b), and a 
double-component model that finally does an adequate job in fitting the 
galaxy (c).  The bottom row shows two more complicated cases (d), which are 
poorly fit even with two components (e), in the end requiring a model with 
six components for the larger galaxy and three components for the smaller 
system (f).
\label{figure 1}}
\end{figure*}
\noindent
portant role of infalling gas in massive halos at high redshift: gas 
accretes directly onto the central galaxy located at the high-density, cooling 
center of X-ray-emitting halo, where stars can form rapidly. 
For the cannibalism scenario, the BCG mainly grows by swallowing 
its surrounding companions as they gradually fall into the central 
system under the influence of dynamical friction (White 1976; Ostriker \& 
Hausman 1977).  Modern scenarios for the formation of BCGs place greater 
emphasis on the role of galaxy mergers (e.g., Dubinski 1998) at different times 
(e.g., Khochfar \& Silk 2006).

Numerical simulations and semi-analytic models suggest that the majority of the 
stars in nearby BCGs were formed in small systems very early ($z\approx2-5$), 
later assembling into a larger entity via a series of mostly dry mergers 
(Boylan-Kolchin et al. 2006; De~Lucia \& Blaizot 2007).  Although 
major dry mergers are still important for the evolutionary history of BCGs 
(Bernardi et al. 2011a, b; Brough et al. 2011), the majority of the merging 
events should have moderate to large mass ratios (Bernardi 2009; Edwards \& 
Patton 2012).   Despite this expectation, most studies have focused on the 
evidence of major mergers in BCGs (Nipoti et al. 2003; Liu et al. 2009). Little 
attention has been devoted to constraining minor mergers in these systems.
 
Ever since their discovery (Malin 1979), shell structures have been considered 
important indicators of merger events in galaxies.  Shells appear as faint, 
arc-like stellar structures that are located concentrically within the main 
body of the galaxy.  Observational evidence of their existence has been 
documented in a number of galaxies (e.g., Schweizer 1980, 1983; Malin \& Carter 
1980, 1983; Carter 1985; Fort et al. 1986; Forbes \& Thomson 1992; Balcells 
1997; Canalizo et al. 2007; Sikkema et al. 2007). Most shells are found in 
galaxies located in low-density environments (Malin \& Carter 1983; Colbert et 
al. 2001). Schweizer (1983) first suggested that shell structures are 
formed by galaxy mergers. This hypothesis was confirmed through numerical 
simulations, which showed that shell structures can indeed be produced in minor 
mergers between an early-type galaxy and a galaxy with much smaller mass 
(Quinn 1984; Dupraz \& Combes 1986; Hernquist \& Quinn 1988, 1989; Cooper et al. 
2011) and major mergers between equal-mass disk-like galaxies (Hernquist \& 
Spergel 1992).  In a minor merger scenario where a massive early-type galaxy 
and a low-mass galaxy have a nearly radial collision, the stars of the smaller 
member are stripped off and oscillate around the larger galaxy under the 
influence of its gravitational potential.  Stars with lower binding energy slow 
down and accumulate at the turning point located at large radii from the center 
of the main galaxy, followed by the formation of inner shells with higher 
energy stars.  All shells move slowly to larger radii after their formation 
(e.g., Quinn 1984; Cooper et al. 2011). As their for-
\vskip 0.1cm  
\begin{figure*}[tb]
\centerline{\psfig{file=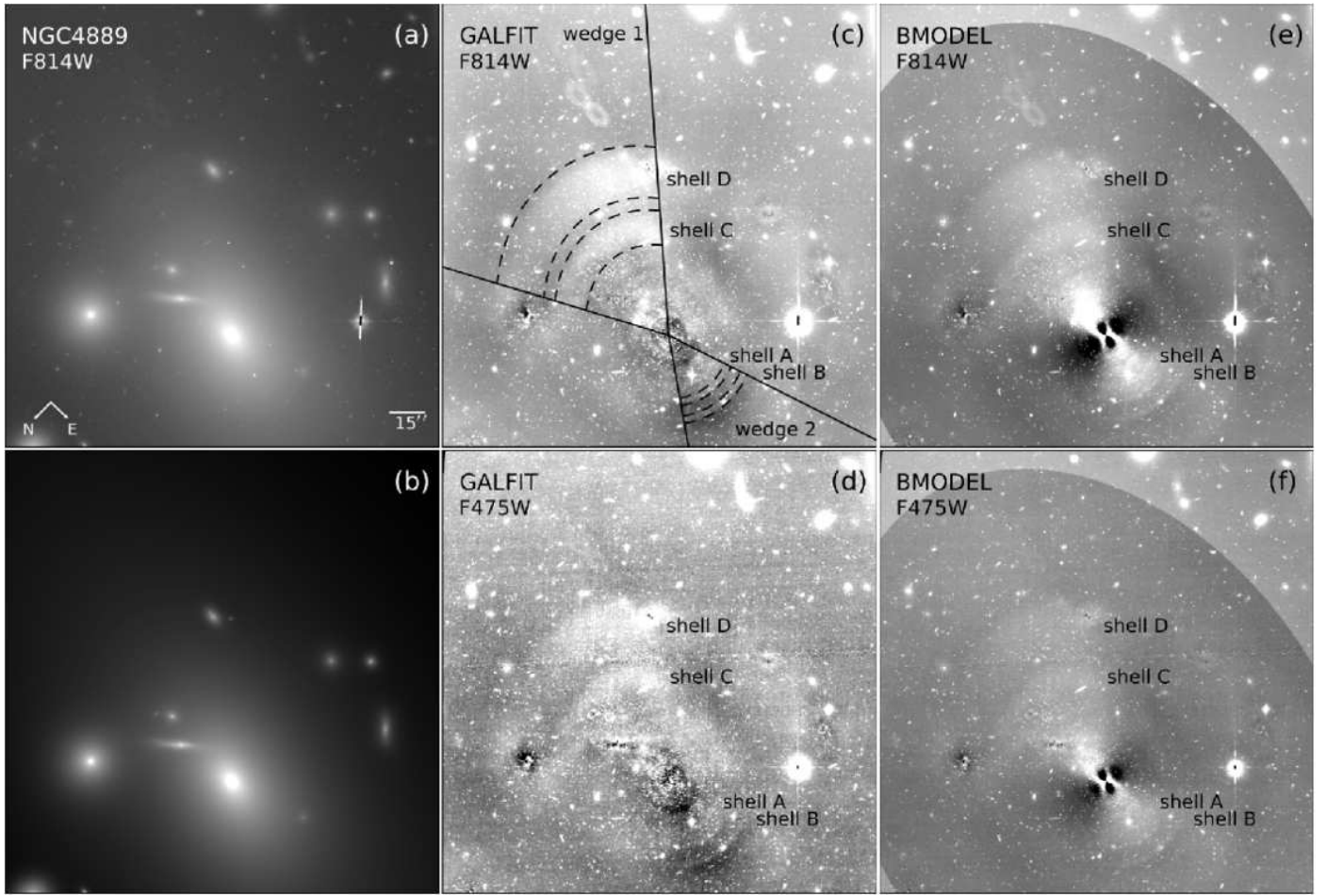,width=18.5cm}}
\figcaption[fig2.eps]{
Overview of the \hst/ACS images and the 2-D analysis.
(a) F814W image of NGC~4889 and its surrounding galaxies.
(b) \galfit\ model of the F814W image.
(c) Residuals from the \galfit\ model of the F814W image; the boundaries of 
wedge 1 and wedge 2 are labeled, as are the locations of shells A--D.
(d) Residuals from the \galfit\ model of the F475W image.
(e) Residuals from the BMODEL of the F814W image. 
(f) Residuals from the BMODEL of the F475W image.
\label{figure 2}}
\end{figure*}
\noindent 
mation mechanism is relatively well understood, 
shells can be used as a tool to diagnose the merger history of massive galaxy. 

Shell structures have been rarely studied in BCGs.  This paper presents the 
discovery of shell structures in the Coma cluster BCG NGC~4889.  We provide 
color measurements that allow us to constrain the nature of the merger that 
produced the shells.  We conclude that NGC~4889 experienced a minor merger with 
a mass ratio of 90 to 1.  

The distance of NGC~4889 is assumed to be 100.0 Mpc, adopted from 
Liu \& Graham (2001). Assuming $H_0$ = 71~km~s$^{-1}$Mpc$^{-1}$, 
${\Omega}_m=0.27$, and ${\Omega}_{\Lambda}=0.73$, the distance corresponds to a 
scale of 0.463 kpc arcsec$^{-1}$.

\section{Data Reduction}

Being one of the BCGs in the Coma cluster, NGC~4889 has been observed by 
various instruments.  For the purposes of analyzing the photometric properties 
of the faint substructures in this massive galaxy, we make use of 
high-resolution optical images from the Advanced Camera for Surveys (ACS) on 
the {\it Hubble Space Telescope} ({\it HST}) and  large field-of-view (FoV) 
optical images from the Sloan Digital Sky Survey (SDSS)
\footnote{http://data.sdss3.org/mosaics}.

As part of Cycle 17 program 11711 (PI: J. P. Blakeslee), NGC~4889 was observed 
by {\it HST}/ACS on March 2010\ in both the F475W (pivot wavelength = 4746.9 
\AA) and F814W (pivot wavelength = 8057.0 \AA) filters. The ACS/WFC detector 
contains two CCD chips, each of size $2048\times4096$ pixels. ACS images provide 
us high spatial resolution and a moderate FoV (202\asec$\times$202\asec) of the 
inner region of NGC~4889. The center of the galaxy is located close to the 
center of chip 2. The western part of the galaxy is better covered, extending to
a radius of 50 kpc. We retrieved two datasets in F475W and one in F814W from the
Mikulski Archive for Space Telescopes (MAST)
\footnote{http://archive.stsci.edu/hst/}. The images, pipeline processed by
{\tt calasc}, have been flat-fielded, subtracted for dark current and bias
level, and corrected for charge transfer efficiency. Each individual dithered 
image has an exposure time of 730~s in F475W and 830~s in F814W. The total
on-source integration time of the final images is 4770~s in F475W and 9960~s in
F814W.

The final ACS images in both filters are generated using the DrizzlePac 
\footnote{http://www.stsci.edu/hst/HST\_overview/drizzlepac} task AstroDrizzle, 
which is an enhanced version of MultiDrizzle (Koekemoer et al. 2002). This
procedure combines the individual dithered images, corrects geometric 
distortion, and rejects cosmic rays. During the AstroDrizzle process, sky 
subtraction is turned on to ensure that each individual frame has the same
pedestal when being combined, even though the sky background values of the two
final output images are measured later. We use a Gaussian kernel to distribute
the flux onto the output images. A series of tests indicates that the Gaussian
kernel does the best job at dealing with bad pixels.  The output pixel scale is 
set to 0\farcs05. During this procedure, the pixels of the multiple input images 
are shrunk by a linear factor {\it pixfrac} before they are mapped to the output 
image. The relative weight of each pixel in the output image is recorded in the
weight image, which is part of the AstroDrizzle final product. Following the
recommendations from the DrizzlePac Handbook (Gonzaga et al. 2012), we use the 
median value and the standard deviation of the weight image to estimate whether
the output image is uniformly covered by the input pixels. To get a decent 
output image, the selected {\it pixfrac} value should allow the standard
deviation of the weight image to be less than 20\% of the median value. We set
the {\it pixfrac} value to 0.8 for F475W and 0.6 for F814W, which results in a
standard deviation of {17.10\%} and {11.79\%} of the median value, respectively, 
for the central $3000 \times 3000$ pixels of the weight images. To ensure that
the images are properly aligned, we set the astrometric center and size of the 
output frames to be the same in the two filters.

We adopted the AB mag system of Oke (1964) and use the updated magnitude zero
points for ACS/WFC: 26.056 mag for F475W and 25.947 mag for F814W
\footnote{http://www.stsci.edu/hst/acs/analysis/zeropoints/zpt.py}.  The
Galactic extinction for NGC~4889 is $A_{\rm F475W}=0.031$ mag and 
$A_{\rm F814W}=0.015$ mag (Schlafly \& Finkbeiner 2011).

\section{Model Fitting}

We construct model galaxies using \galfit, a fitting program 
that uses multiple analytic functions to perform two-dimensional (2-D) 
decomposition of galaxy images (Peng et al. 2002, 2010).  We adopt the \ser\ 
(1968) function, which has enough flexibility to describe the surface 
brightness distribution of various types of galaxies. 
Our primary goal is to detect faint substructures in NGC~4889.  
To achieve this, we build a monotonically smooth model of the underlying 
galaxy using as many components as necessary to achieve the best 
global fit, and then we subtract the best-fit model from the original 
data to derive a residual image that accentuates low-contrast, low-surface 
brightness substructure.  We ascribe no physical significance to each 
individual component or to the total number of components.  \galfit\ normally 
needs an accurate model of the point-spread function to recover the intrinsic 
value of the parameters for each model component.  To save computation time, 
we do not account for point-spread function convolution.  This has no 
consequence on our results because we do not attempt to extract physical 
information from the model parameters.  Moreover, the low-surface brightness 
features we seek are located far from the center of the galaxy; an accurate 
model for the central regions of the BCG is not necessary for our purposes.  

\galfit\ uses ${\chi}^2$ statistics to minimize the difference 
between the image and the model; hence, the reduced ${\chi}^2$ value should, 
in principle, be the natural tool to evaluate the goodness of the fit.  
However, for large images such as the ones used in our work, it is challenging
to assess the robustness of the individual pixel uncertainties required for 
the calculation of ${\chi}^2$.  Moreover, the reduced ${\chi}^2$ is a single 
parameter that can provide only a very general impression of the quality of 
the fit.  For our
\vskip 0.2cm
\centerline{\psfig{file=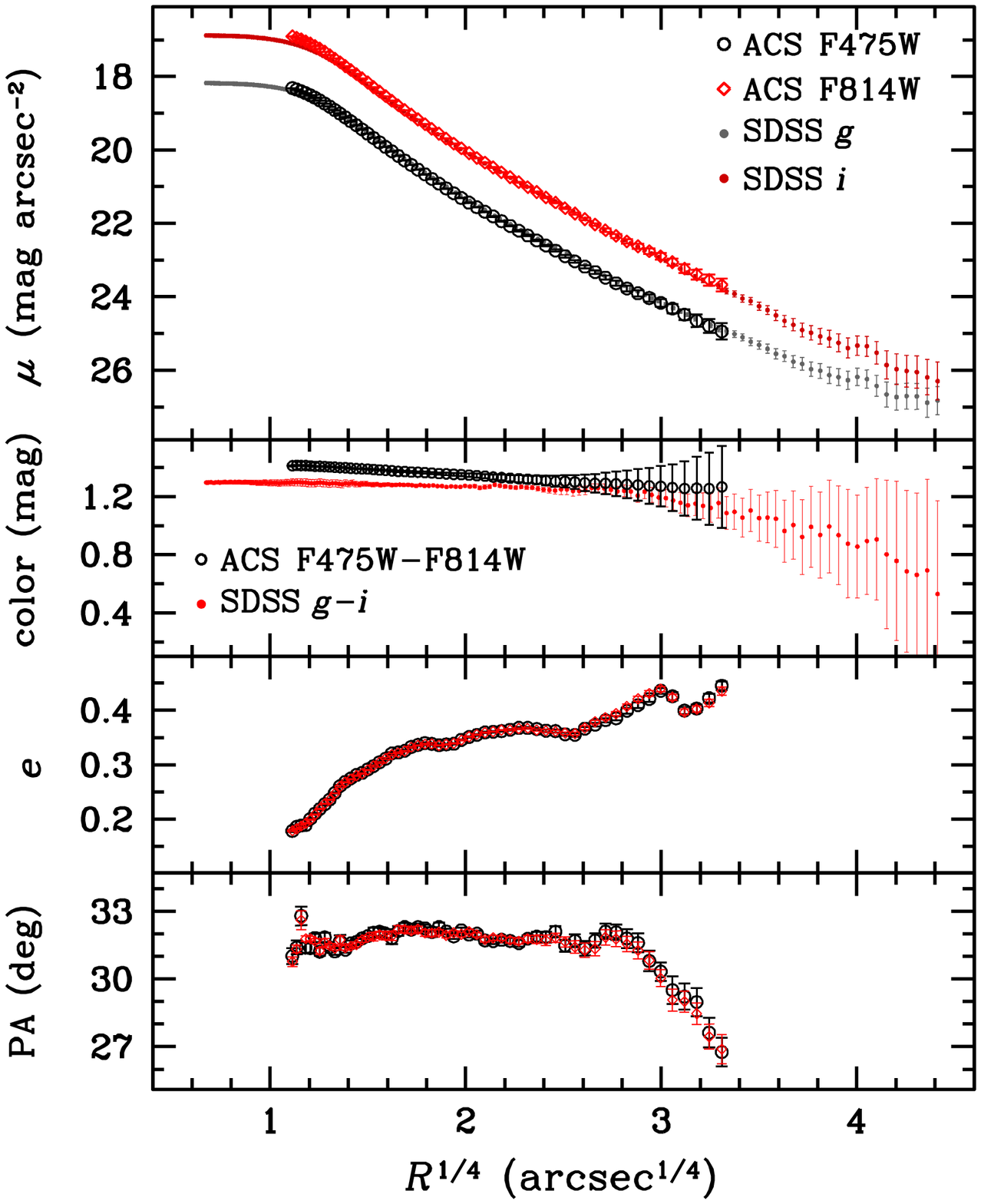,width=8.75cm}}
\figcaption[fig3.eps]{
1-D isophotal analysis of \hst/ACS and SDSS images of
NGC~4889. From top to bottom, we show the radial profiles of surface
brightness, color, ellipticity, and position angle.
\label{figure 3}}
\vskip 0.4cm
\noindent
current application, the most straightforward and reliable 
metric for judging the goodness of the fit is simply from visual inspection 
of the model-subtracted residual image.  The best model is that which achieves 
the smoothest, flattest residuals using the minimum number of components.  
This strategy closely resembles that used by Huang et al. (2013) to  study the 
multi-component nature of nearby elliptical galaxies.

We first fit NGC~4889 and its surrounding smaller galaxies with 
single-component S\'ersic models, and then we gradually build up the 
complexity of the models by adding additional components, if warranted.  After 
each trial, we visually evaluate the quality of the fit and the parameters of 
the model components.  If we encounter any redundant component (one with 
properties very similar to those of another component) or a component with 
unreasonable parameters (e.g., very large effective radius, very large 
ellipticity, or unusually faint central brightness) that does not lead to 
noticeable improvement in the residual image, we stop the process even though 
additional components may lead to a lower reduced ${\chi}^2$ value.  Figure 1 
illustrates our method, showing two examples of relatively small galaxies near
NGC~4889.  For the relative simple case in the upper row, a model with 
two \ser\ components is sufficient because the residual image is already very 
smooth.  On the other hand, the two galaxies on the bottom row require more 
complicated models to achieve a residual image of the same quality.  
Six components are needed for the larger galaxy on the left and three for the 
smaller system on the right.

\vskip 0.2cm
\centerline{\psfig{file=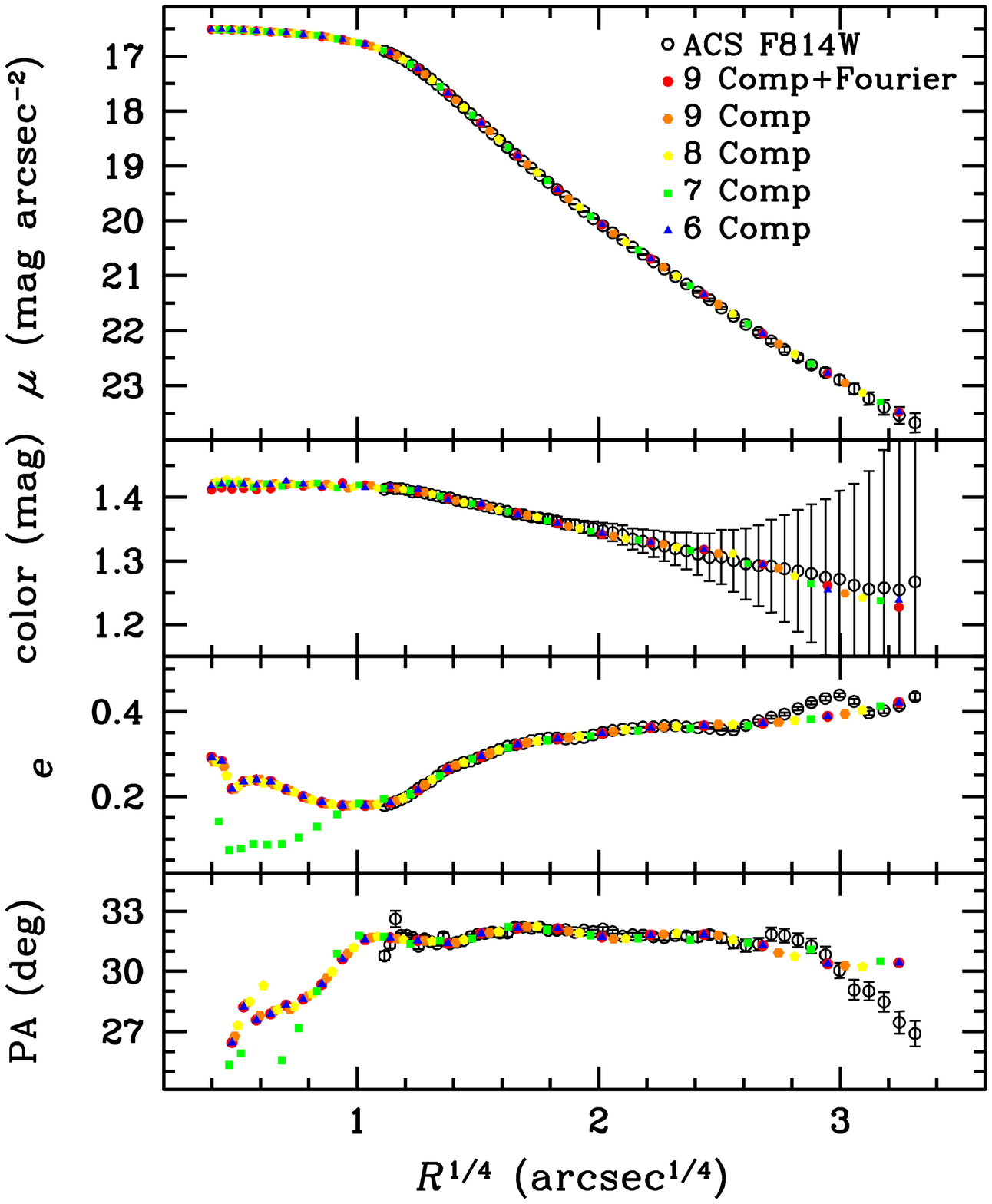,width=8.75cm}}
\figcaption[fig4.eps]{
1-D isophotal analysis of the \hst/ACS F814W image and all \galfit\ models of
NGC~4889.  From top to bottom, we show the radial profiles of surface
brightness, color, ellipticity, and position angle.  Different models are
represented by symbols of different shapes and colors.
\label{Figure 4}}
\vskip 0.4cm
\noindent

\subsection{Subtraction of Surrounding Galaxies}

As the BCG is embedded in a crowded field (Figure~2), we also need to 
simultaneously fit and remove all the neighboring galaxies.  In total we model 
14 small galaxies surrounding NGC~4889.  When fitting each of them, we select 
a rectangular patch covering the target and its nearby region, and mask out 
irrelevant sources on the patch. The aim is to describe the small galaxies
in as much detail as necessary to remove them.  We model each galaxy with
multiple \ser\ components, enabling Fourier modes (Peng et al. 2010) as needed. 
There are two disk-like galaxies. To model the smaller one 
(SDSS~J13006.11+275841.9) located to the northeast of NGC~4889, we used nine 
non-concentric components, adding first, third, and fourth Fourier modes to the 
seven largest components. For the larger disk galaxy (2MASX~J13001036+2757332) 
located to the west of NGC~4889, we added fourth and sixth Fourier modes to all 
of its four concentric components. The remaining galaxies were fit well without 
Fourier modes. The sky background of the neighboring galaxies is the extended 
envelope of NGC~4889, which has noticeable local curvature due to the BCG.  
We approximate the background curvature using a large \ser\ component that is 
centered somewhere outside the image.

We start with the high--signal-to-noise ratio F814W image. Since each small 
galaxy may affect the model construction of its surrounding counterparts, we 
adopt the following iterative procedure. We first construct a preliminary 
model for NGC~4889 and the surrounding galaxies. Then we refit each galaxy 
using the residual background obtained after subtracting all the other galaxies,
thereby producing a refined model for each galaxy.  We repeat this procedure 
until the overall residual pattern is visually satisfactory. Even after 
we model and subtract the 14 brightest neighbors, the residual 
images reveal several additional small galaxies projected close to the center 
of NGC~4889.  We do not model these further, and instead simply mask them out 
during further analysis.

To ensure that we can obtain meaningful color measurements, we must constrain 
the models constructed from the two filters.  The model parameters for the F475W 
image are tied to those of the F814W image. We fix all the parameters except the
absolute positions and integrated magnitudes, to allow slight misalignment and
color difference between the two images. The relative positions of the
subcomponents in each model are, however, constrained to each other. 

\subsection{Sky Background Determination}

The ACS images do not cover sufficient area to enable a reliable measurement
of the sky background, which is essential for our photometric analysis.  We 
estimate the background of the ACS images by scaling the surface brightness 
profile of the F475W image to that of a wide-field SDSS $g$-band (pivot 
wavelength $=4770$ \AA) image over a range of radii unaffected by their 
resolution difference.  The same procedure is used to match the F814W image 
with the SDSS $i$-band (pivot wavelength $=7625$ \AA) image.  The large-FoV 
ground-based mosaic images were generated from the Imaging Mosaic tool of
SDSS Science Archive Server (SAS).  They have already been corrected for 
the sky by the method described in Blanton et al. (2011); each has a field of 
0.65\deg$\times$0.65\deg, a pixel scale of 0\farcs396, and is centered on  
NGC~4889.

We use the {\tt IRAF}\footnote{STSDAS is a product of the Space Telescope 
Science Institute, which is operated by AURA for NASA} task ELLIPSE 
(Jedrzejewski 1987) to derive one-dimensional (1-D), azimuthally averaged
intensity profiles along the semi-major axis of NGC~4889 on the ACS images, 
after first subtracting the neighboring contaminating galaxies.  We apply a 
mask image that contains three components: (1) the bad pixel map obtained from 
the weight image as part of the AstroDrizzle output; (2) bright surrounding 
sources detected by SExtractor (Bertin \& Arnouts 1996); and (3), to minimize 
the effect of the uneven central residual pattern on the isophotal analysis, a 
manually generated mask for the central region of each surrounding galaxy 
on the model-subtracted image.

We first run ELLIPSE on the ACS images with fixed central position, which we 
determined from \galfit\ earlier. The ellipticity and position angle are 
allowed to vary.  The radially averaged values of the ellipticity and position 
angle are then fixed during the next execution of ELLIPSE applied to the mosaic
SDSS images.  To determine the sky level, we generate an aggressive mask to 
cover all the bright sources. We median-smooth the images using boxes of 
different sizes ranging from 3$\times$3 to 55$\times$55 pixels. The unmasked 
pixel value distribution of each smoothed image is extracted after a $4\sigma$ 
clipping algorithm is iterated for 6 times to reject outliers.  We then fit 
this distribution with a Gaussian function, whose peak gives the expected sky 
level and the width is taken as the uncertainty. From the relation between the 
box size for median-smoothing and the uncertainty value, we find that the 
uncertainty first increases with the box size but remains very stable after 
$20\times20$ pixels, which reflects the intrinsic background fluctuation level.  
In the end,
\vskip 0.1cm
\begin{figure*}[t]
\centerline{\psfig{file=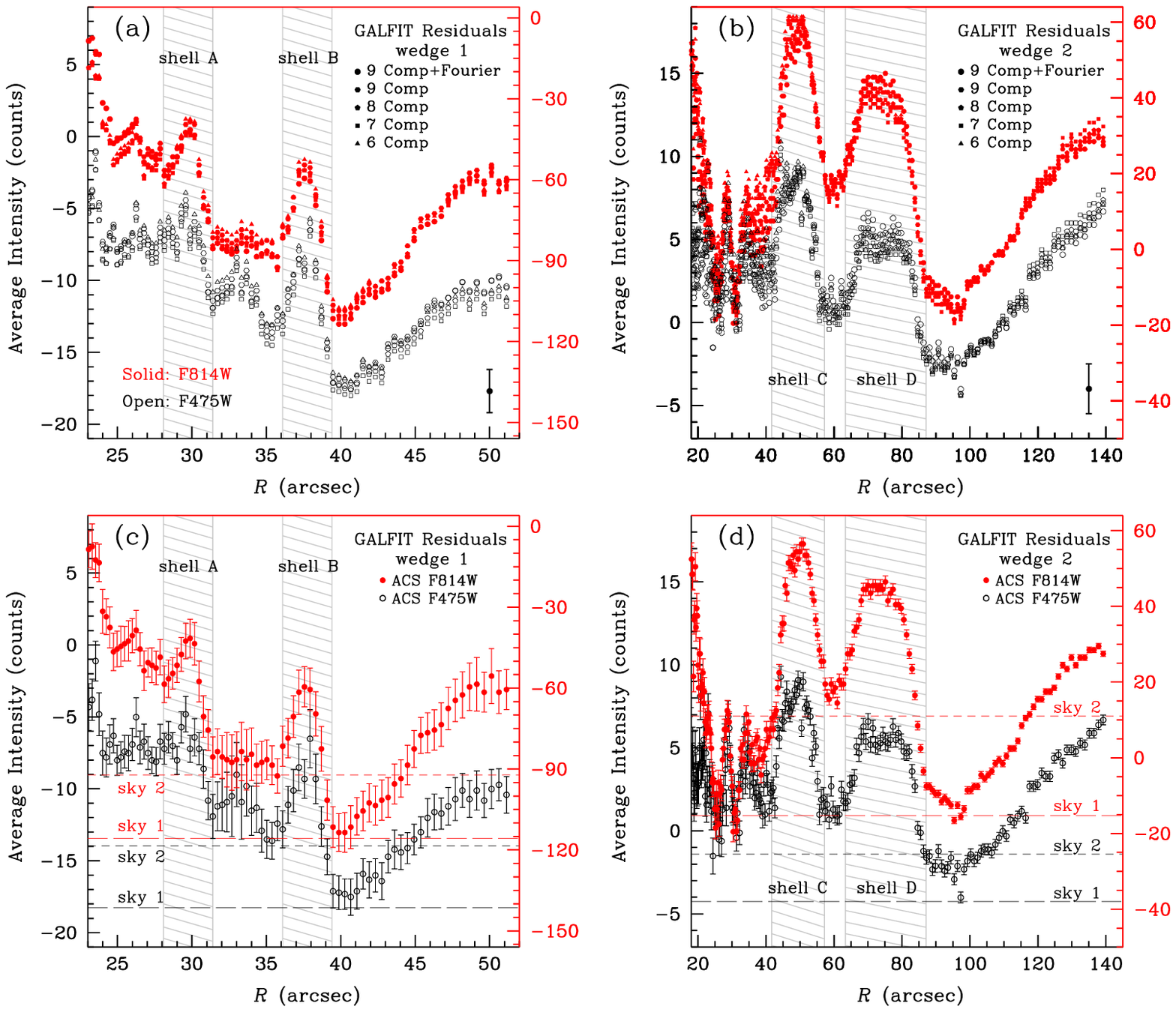,width=14.75cm}}
\figcaption[fig5.eps]{
Average 1-D intensity profiles of the shell
structures generated by subtracting the best {\tt GALFIT} model from the
original images. The shells in each wedge are defined within the radii shaded
by the hatched lines.  The profiles for the F475W and F814W bands are shown
as red (right ordinate) and black dots (left ordinate), respectively. In
panels (a) and (b), the results from different models are represented by
different symbols. A typical error bar is shown on the lower-right corner.
Panels (c) and (d) give the intensity profiles generated from the final,
best-fit model, which consists of of nine S\'{e}rsic components with Fourier
modes enabled.  The red and black horizontal dashed lines show the level of
sky 1 and sky 2 for the F475W and F814W bands, respectively.
\label{fig5}}
\end{figure*}
\vskip 0.15cm
\noindent
we adopt 
the uncertainty values from images smoothed by a $20\times20$ box: the value 
is $(1.50\pm12.7)\times10^{-4}$ counts~s$^{-1}$ for $g$ and 
$(8.3\pm33.0)\times10^{-4}$ counts~s$^{-1}$ for $i$.

Next, we derive the $g$-band and $i$-band surface brightness profiles of the 
SDSS images, after applying object masks generated from SExtractor.  The 
ellipticity and position angle are fixed to the values determined from the 
ACS images.  Because of the coarse resolution of the SDSS images,  
a larger step size (0.05 compared with 0.02) is adopted for incrementing the 
factor by which the semi-axis length is increased between successive 
isophotes.  We estimate the background level of the ACS images by matching the 
F475W and $g$ profiles and the F814W and $i$ profiles (Figure~3).  We neglect 
the central 3\asec\ region to avoid the obvious mismatch between the 
{\it HST}\ and SDSS point-spread functions.  On larger scales the comparison 
can only be made out to $\sim$115\asec\ because the ACS images have a more 
limited FoV compared with the SDSS.  We use a non-linear least-squares 
optimization method to join up the brightness profiles. During 
the fitting procedure, two variables are optimized simultaneously. The 
first variable is the sky background value of the ACS images. The second 
variable is a constant offset in the overall photometric zero point that 
practically accounts for the difference in photometric zero point between the 
ACS and its corresponding SDSS filter.

The intrinsic fluctuation of the background level of the ACS 
images is an important factor that affects the uncertainty on the sky.  We 
first explain here the procedure for measuring the intrinsic fluctuation of the 
background and defer the discussion on model 
construction and shell detection for the next section.  We measure the 
intrinsic fluctuation on the \galfit\ residual images after subtracting the 
surrounding galaxies and the model for NGC~4889 from the original data. We 
apply the mask used in constructing the \galfit\ models to exclude all the 
bright sources and faint shell structures. Then we apply a procedure 
similar to that used for the SDSS images: we median-smooth the images using 
boxes of different sizes, ranging from 3$\times$3 to 55$\times$55 pixels, 
and we apply a $4\sigma$ clipping algorithm 6 times to reject outliers on 
the smoothed images. A Gaussian function is then fit to the unmasked pixel; 
the width of the distribution is taken as 
the uncertainty. Adopting, as before, the images smoothed by a $20\times20$ 
box, the intrinsic fluctuations are $1.326\times10^{-3}$ counts~s$^{-1}$ for 
the F475W image and $3.002\times10^{-3}$ counts~s$^{-1}$ for the F814W image. 
The final uncertainty of the sky level of the ACS images is the combination of 
the sky uncertainty of the SDSS images, the uncertainty from the least-squares 
optimization method, and the intrinsic fluctuation of the ACS images.
\begin{figure*}[t]
\centerline{\psfig{file=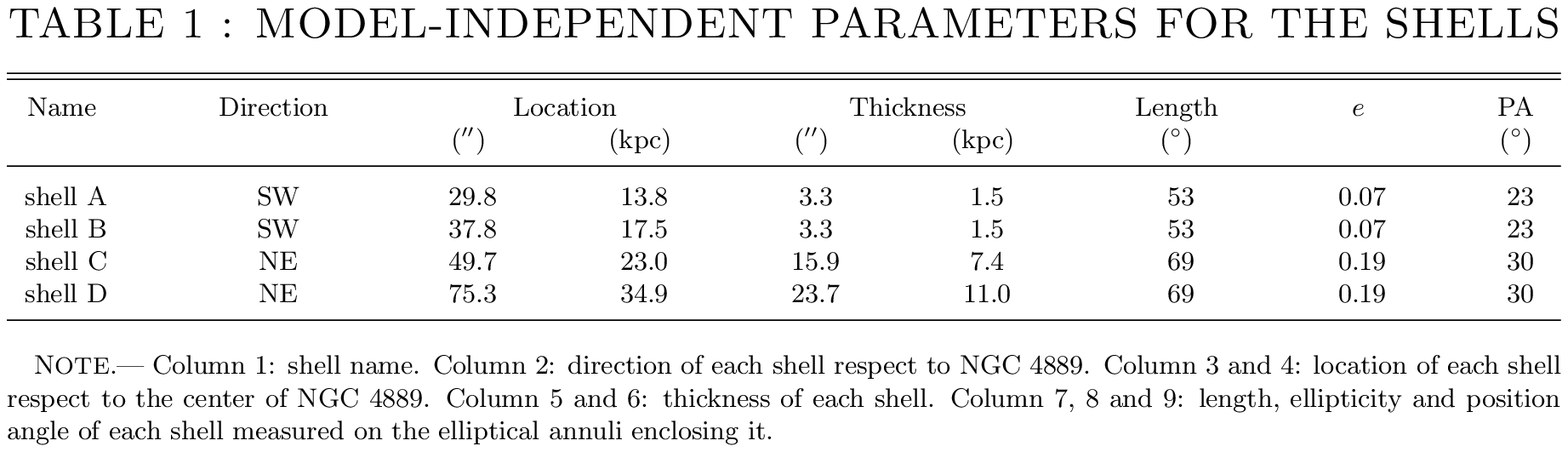,width=16.5cm,angle=0}}
\end{figure*}
\noindent

\subsection{Detection of Shell Features in NGC~4889}

Shells, as low-surface brightness features, are difficult to detect and measure
in the presence of their bright host galaxies. To reveal the photometric 
properties of shell structures, we use the residual images made by subtracting 
the model of the host galaxy.  Because shells are very faint, even small
deviations in the host galaxy model can greatly affect the inferred light 
distribution of the shell structures on residual images. The photometric 
properties of shell structures are therefore strongly model-dependent.  We 
require an accurate and carefully constructed model for the host galaxy. We 
build a host galaxy model two ways, first by using {\tt GALFIT} and then 
independently verifying it using a non-parametric approach based on isophotal 
analysis.  In both cases we first subtract the surrounding small galaxies from 
the images.

We built five different host galaxy models using \galfit. Each model consists 
of three to nine \ser\ components and a sky component.  During the fitting we 
apply a mask generated from the combination of the mask used for ELLIPSE plus an
additional one that covers all the shell regions.  With the shell regions masked
out, we are able to build models that describe only the host galaxy without the
influence of the shell structures. The average sky background is fixed to the 
value derived in Section 3.2.  In view of possible variations in the local
background due to the nearby galaxies or intracluster light, we further allow 
the sky gradient to be adjusted.  The fits are first performed on the deeper
F814W image, with all the parameters for each \ser\ component free to vary. 
Most of the resulting parameters are then used and fixed in the F475W model, 
including the effective radius, \ser\ index, ellipticity, and position angle. 
Only the integrated magnitude of each component is allowed to vary.  Models 
generated in F475W and F814W have exactly the same size and shape.

An initial three-component fit of NGC~4889 suffices to reveal the presence of 
shell structures, but strong features remain in the residual images.  Six 
components give a much flatter residual pattern, but the innermost part of the 
galaxy, mostly within 200 pixels from the center, is still not well fit.  We 
gradually increased the number of components, one at a time, eventually finding 
the optimal solution to be nine components with the 4th and 6th Fourier modes 
turned on for the six innermost components.  The middle panels of Figure~2 give
the residual images from our final, best-fit nine-component model with Fourier 
modes enabled.   For completeness, however, we illustrate the other four
acceptable models (Figure~4).  All show satisfactory 2-D residual patterns and 
rather similar 1-D surface brightness, ellipticity, and position angle profiles.

We use a non-parametric approach to independently confirm the reality of the 
shell structures and that they are not artifacts generated by our parametric 
fits. We use the {\tt IRAF} task BMODEL to construct a smooth model of NGC~4889 
using the best-fitting isophotes generated from the 1-D analysis using ELLIPSE.  
We manually correct the sky background using the sky level found in Section~3.2, 
and we run ELLIPSE with the ellipticity and position angle fixed to the values 
derived in the previous section.  To trace the galaxy smoothly and with the best 
fidelity, we set the fitting step to 0.05 and compute the median rather than the
mean intensity.  The BMODEL residual images are shown on the right column of
Figure~2.  It is reassuring that the overall shell features are robust 
substructures: their general appearance is very similar in the \galfit\ and
BMODEL residual images.

Despite the overall similarity of both methods in revealing the presence of 
shells, we advocate that the \galfit-based 2-D method as the better choice for 
quantitative analysis of the shell features, for the following reasons.  The 
\galfit\ approach allows us to construct consistent, well-constrained models
for images taken in different filters (e.g., by coupling the geometric 
parameters of the various subcomponents).  This is absolutely crucial to derive 
robust colors for the shell features.  Moreover, \galfit\ has better ability to 
handle images that have been aggressively masked, as is necessary for the very
complicated environment surrounding NGC~4889 and other BCGs. By contrast, 
ELLIPSE and other isophote fitting routines will often fail when confronted with 
such large regions of flagged pixels.  \galfit\ uses the information in the 
entire 2-D image, while ELLIPSE fits each isophote using only the information 
around a pre-defined semi-major axis length. Even small local irregularities in
the image that are not perfectly masked can influence the mean intensity of the
corresponding elliptical isophote. A 2-D reconstruction of the isophotes using 
BMODEL produces ring-like structures in the residual images that can be mistaken 
for real shell features.  

\section{Location and Color of the Shells}

Four shells are unambiguously detected in the residual images 
generated from both the \galfit\ and BMODEL models. We highlight the shells 
with dashed lines on Figure~2. Ordered according to their distance 
from the center of NGC~4889, they are marked as shells A (13.8 kpc), 
B (17.5 kpc), C (23.0 kpc), and D (34.9 kpc). Shells A and B are located to the 
northeast of the galaxy, and shells C and D are located on the southwestern 
side. At first glance, all the shells are roughly aligned with the major axis 
of the host galaxy. The position angle is 23\deg\ for shells A and B and 30\deg\ 
for shells C and D, to be compared with a position angle of 31\deg\ for the host 
galaxy.  The inner two shells are much thinner (3\farcs3) and shorter (53\deg) 
than the outer 
\begin{figure*}[t]
\centerline{\psfig{file=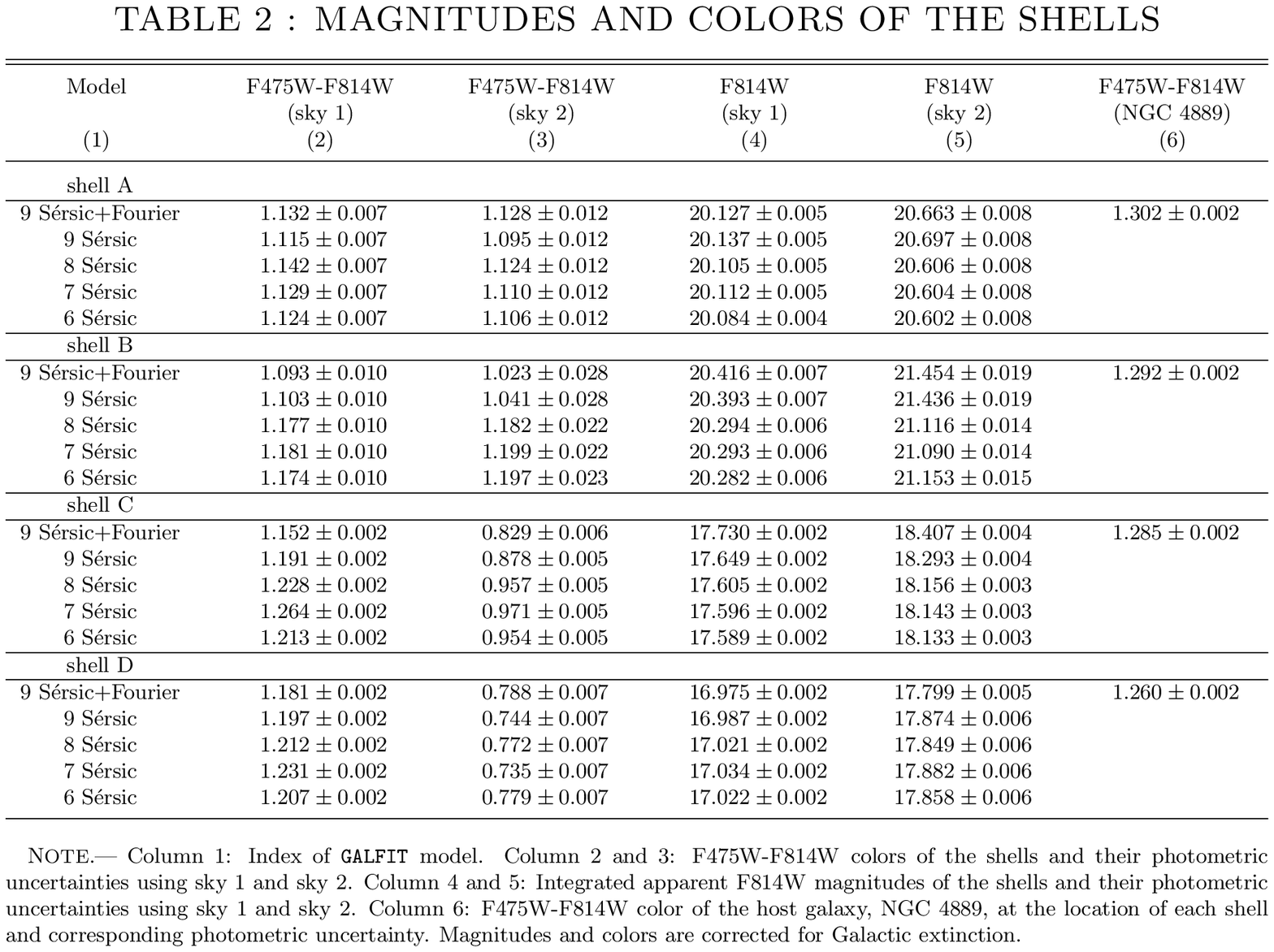,width=16.5cm,angle=0}}
\end{figure*}
\noindent
two shells (15\farcs9 and 69\deg\ for shell C; 23\farcs7 and 
69\deg\ for shell D). By comparing the residual images generated using the two 
methods, it is obvious that \galfit\ gives a more distinct view of the shell 
structures.  The residuals from BMODEL have fan-like artifacts that are not 
present in the \galfit\ residuals.  

We first need to define the physical boundaries of the shells in order to be 
able to measure their photometric properties.  We obtain the basic geometric 
parameters for the four shells by visually drawing elliptical annuli that 
approximately enclose the shell-like brightness enhancements visible on the 
F814W residual image.  Table~1 lists the derived radii, thickness, length,
ellipticity, and position angle of the shells.  Based on the geometric 
information of the shells, we establish two wedge-shaped regions to enclose 
them, as shown on Figure~2c.  Wedge~1 encloses shells A  and B; Wedge~2 contains
shells C and D.  We then use ELLIPSE to calculate the average intensity profile
of each shell within the wedges. The center of the fitted isophotes are all 
fixed at the center of NGC~4889, and the ellipticity and position angle are 
fixed to the values determined for each shell.  

Figures~5a and 5b show the average intensity profiles for the shells.  Results 
from the different \galfit\ models are given.  While the intensity profile is 
very complex, we can clearly identify four peaks that correspond to the four 
shells seen in the 2-D residual images.  The location of the peaks are marked 
with light shaded regions.  Overall there is good agreement among the different
models.  The basic shapes of the shell profiles are robust, and the main 
differences come from slight offsets in the normalization of the intensity.  
There is also good correspondence between the two filters.  Even though the
F475W data set is noisier, it clearly traces the same overall structures seen 
more clearly in the F814W image.

We perform aperture photometry on the \galfit\ residual images to determine 
the integrated apparent magnitude and color of the shells, using apertures 
defined using the geometric parameters established above.  Although the ACS 
images have had their global sky background subtracted, as described 
in Section~3.2, we need to redetermine the {\it local}\ background for each
shell in order to properly measure its brightness because it sits on top of 
significant diffuse emission that presumably belongs to real signal from 
NGC~4889 and/or intracluster light.  We illustrate our approach in Figure~5c and 
5d, using, as an example, the 1-D average intensity profiles generated from the 
best-fitting nine-component \ser\ model with Fourier modes.  Within each wedge 
and for each shell, we define the local background to be one of the two 
low-intensity regions determined from the 1-D profiles.  Within each wedge, 
``sky 1'' is located beyond the two shells, while ``sky 2'' is located between 
the two shells. The sky regions have the same range in radius, ellipticity, and
position angle as those of shells within the wedge. Their inner and outer radii 
are carefully estimated by visual inspection of the 1-D profiles, to ensure that
they enclose all the low-intensity pixels. Within each sky region, we measure 
the sky value using the same method employed for measuring the sky on large 
scales (Section~3.2).  Namely, (1) we mask out any remaining bright sources; (2) 
we apply six iterations of $4\sigma$ clipping; and (3) we fit a Gaussian to the 
distribution of sky pixel values and then adopt the peak of the Gaussian as the 
local sky level and the width of the Gaussian as the error on the sky.  Note 
\vskip 0.1cm
\begin{figure*}[t]
\centerline{\psfig{file=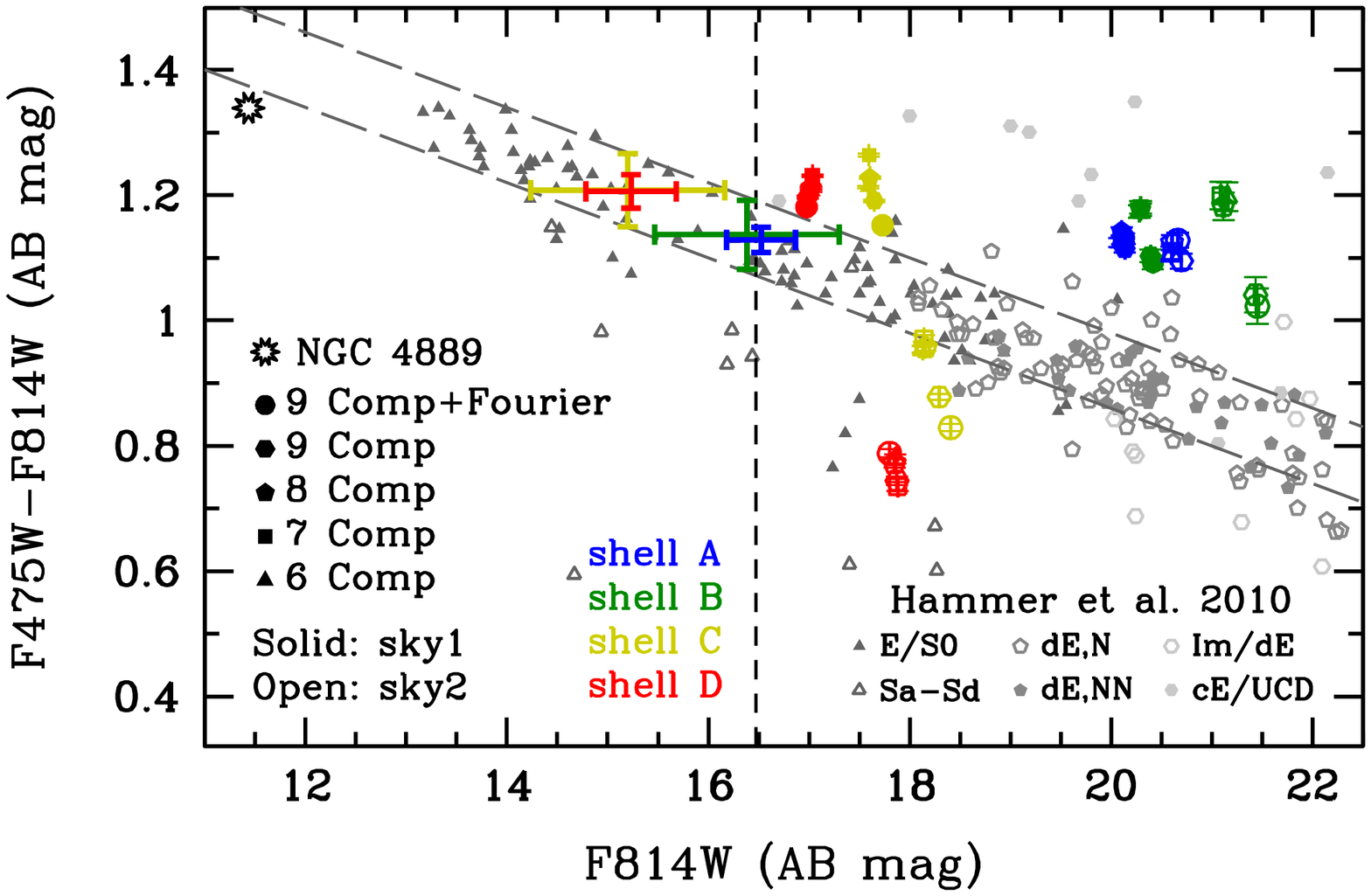,width=16.5cm}}
\figcaption[fig6.eps]{
The CMR of Coma cluster galaxies as given by Hammer
et al. (2010; diagonal long-dashed lines). The gray points represent data from
Hammer et al. (2010). Different galaxy types are marked by different symbols
and slightly different shades of gray. The measured magnitudes and colors of
the four shells are shown in blue, green, yellow, and red symbols,
with the different models denoted by various symbol types. The magnitude of
each shell predicted from the CMR is plotted as error bars in the
corresponding color. NGC~4889 is plotted as a black star. The vertical dashed
line marks the total magnitude of the four shells.
\label{figure 6}}
\end{figure*}
\vskip 0.1cm
\noindent
that, because of sigma-clipping, the actual values of sky 1 and sky 2 are 
slightly lower than the corresponding values inferred from the minima of the 
1-D profiles.

The integrated apparent magnitude of each shell is calculated using the 
following formulae:

\begin{equation}
mag = zmag - 2.5 \, \log \left(\frac{{flux} - A \times{sky}}{t_{\rm exp}}\right)
\end{equation}

\begin{equation}
{\sigma}_{mag} = 1.0857 \, \frac{\sigma}{flux - A \times sky}.
\end{equation}

\noindent
In the above equations, $zmag$ (mag) is the photometric zero 
point for the respective filter, $flux$ (counts) is the total flux within the 
aperture, with associated error $\sigma$, $A$ is the number of pixels in the 
aperture, $sky$ (counts/pixel) is the sky level, and $t_{\rm exp}$ (seconds) is 
the exposure time.  We use both sky 1 and sky 2 for the sky level. We calculate 
${\sigma}$ using 

\begin{equation}
{\sigma} = \sqrt{\frac{flux - A \times sky}{gain} + A \times {\sigma}_
{sky}^2 + \frac{A^2 \times {\sigma}_{sky}^2}{A_{sky}}}.
\end{equation}

\noindent
Here, gain (e-/ADU) is the gain value of the detector.
Because of their low surface brightnesses, the final magnitudes and especially 
the colors are highly model-dependent.  The measurements are sensitive 
not only to the global \galfit\ model adopted, but also to the choice of local 
sky background.  The measurements for  all five \galfit\ models and using both 
prescriptions for local sky determination are presented in Table~2 and 
illustrated in Figure~6.  These measurements bracket the full range of possible
systematic uncertainties.  We believe that the most robust measurements come 
from the nine-component \ser\ model with Fourier modes combined with the 
local background estimated using sky 1.  The apparent magnitude of all four 
shells combined is $m_{\rm F814W} = 16.453 \pm 0.018$; it is marked by the 
vertical dashed line on Figure~6.  The colors for shells A and B are quite
similar for sky 1 and sky 2, but they differ significantly for shells C and D.
Adopting sky 1 as the fiducial background, the color of shells 
A--D are ${\rm F475W}-{\rm F814W} = 1.129 \pm 0.021$, $1.137 \pm 0.055$, 
$1.208 \pm 0.058$, and $1.206 \pm 0.027$i, respectively.  The shells have 
colors in the range F475W$-$F814W $\approx 1.1-1.2$ mag.  The shell colors are
significantly bluer than the integrated color of NGC~4889, which is 
F475W$-$F814W = 1.34 mag within an aperture with a semi-major axis length of 
40\asec, similar to that adopted by Carter et al. (2009).  Moreover, each shell 
is slightly, but measurably {\it bluer}\ than the color of NGC~4889 at the 
location of each shell, which lies in the range F475W$-$F814W 
$\approx 1.26-1.30$ mag.  This represents one of the very few robust color 
measurements of shells in BCGs, and the first to show unambiguously that the 
shells have different colors than the underlying host galaxy.

\section{Physical Implications}

Minor mergers, especially those that involve little or no dissipation, are now 
frequently invoked to explain the mass and size evolution of massive early-type 
galaxies. The accretion of smaller stellar systems without significant cold gas 
can help build up the extended envelopes of massive ellipticals. The shell 
structures we found in NGC~4889 can be regarded as an evidence of the type of 
minor merger event envisioned. The location of NGC~4889 in Coma 
suggests that the shells were probably not generated through gas-rich or major 
mergers: the probability of either is low in the core of a rich cluster.  
Instead, the shells were most likely formed from a minor merger (Quinn 
1984; Dupraz \& Combes 1986; Hernquist \& Quinn 1988, 1989; Cooper et al. 2011).
Furthermore, the lack of dust lanes or other substructures in the residual 
images of NGC~4889 that might indicate the presence of cold gas strongly 
hints that the minor merger event was dry.

Beyond their mere discovery, the shells in NGC~4889 can be used to deduce 
a quantitative estimate of the mass of the accreted galaxy.  The key lies in 
the colors of the shells. 

Early-type galaxies in clusters obey an extraordinary tight color-magnitude 
relation (CMR): more luminous galaxies are redder (Faber 1973; Visvanathan \& 
Sandage 1977).  Although broad-band optical colors are notoriously difficult 
to interpret because of the degeneracy among age, metallicity, and dust 
reddening, the situation is considerably cleaner for early-type galaxies, 
especially those in clusters, which, to zeroth order, are old and dust-free.
The CMR for early-type galaxies is widely viewed as a correlation between 
metallicity and luminosity or stellar mass (Faber 1973; Larson 1975; Kodama \& 
Arimoto 1997). 

The Coma cluster, in particular, exhibits a well-developed, tight CMR covering 
a wide range of galaxy types and luminosities (e.g., Dressler 1980; Godwin et 
al. 1983; Bower et al. 1992; Terlevich et al. 1999, 2001). The recent 
measurements from the \hst/ACS Coma Cluster Survey (Hammer et al. 2010) are 
especially useful for our purposes because it used exactly
the same filter pair we used.  Hammer et al. (2010; their Figure~13) constructed
a deep CMR for Coma that extends down to about 22.0 AB magnitude in the F814W
filter. Their sample is reproduced in Figure~6, where we now add our 
measurements of the four shells in NGC~4889 and the integrated quantities for
NGC~4889 itself. Each shell is coded by a different color, and the results for
the different \galfit\ models are shown with different symbols.  Focusing only 
on the measurements using sky 1, the color of shells A--D are 
${\rm F475W}-{\rm F814W} = 1.129 \pm 0.021$, $1.137 
\pm 0.055$, $1.208 \pm 0.058$, and $1.206 \pm 0.027$, 
respectively, where the error bar represents the dispersion among the five 
models.  The diagonal long-dashed lines bracket the scatter of Hammer et al.'s 
fit to the CMR for Coma:

\begin{equation}
{\rm F475W}-{\rm F814W} = -0.06 \times {\rm F814W} + 2.12.
\end{equation}

\noindent
It is immediately clear that the shells lie significantly and systematically 
{\it above}\ the CMR.  The individual shells are ``too red'' for 
their luminosities.  This is true for all of the shells if the colors are 
derived using sky 1 as the background, and it is true of shells A and B even if 
we assumed sky 2.  As explained in Section~4, we think that sky 1 is more 
trustworthy than sky 2, and for the rest of the paper we will focus only on 
measurements made using sky 1.  

We can naturally explain the anomalous position of the shells under the 
hypothesis that the shells represent the trace remnants of a once-disrupted 
larger galaxy.  The shells bear the imprint of the metallicity (or color) of 
their progenitor galaxy, {\it before}\ it fell into NGC~4889 and was torn 
apart.  We assume, reasonably, that the progenitor was a member of Coma and 
that it used to follow the CMR of the cluster.  We further assume that all 
four shells were formed in a single merger event, and that on average the 
colors of the shells are similar to the color of the progenitor galaxy.  By 
extrapolating the observed colors of the shells back to the CMR, we can 
constrain the original luminosity (or mass) of the progenitor galaxy.
 
The large colored error bars show the predicted locations of the progenitor on 
the CMR.  The magnitudes range from as bright as F814W = 15.17 for shells C 
and D to as faint as 16.50 for shell A.  For an integrated magnitude of F814W 
= 11.44 for NGC~4889, measured within the same 40\asec\ aperture used to 
obtain its colors, the range of possible progenitor luminosity (mass) ratios 
(with respect to NGC~4889) is 31:1 to 106:1.  According to numerical simulations 
(e.g., Quinn 1984), the innermost shells form last.  We thus expect the inner 
shells to better reflect the actual properties of the progenitor galaxy as they 
should be less contaminated by mixing with the ambient stars.  Within our data 
set, the colors of the two inner shells have the additional advantage of being 
better measured than those for the two outer ones.  Shell B, in particular, is
relatively isolated and has the best determined local background (see Figure~5). 
With ${\rm F475W}-{\rm F814W} = 1.137 \pm 0.055$ mag, the CMR 
predicts F814W = $16.33$ mag for the progenitor, which corresponds to a 
luminosity (mass) ratio of 90:1.  The observed total magnitude of the four 
shells, F814W = $16.453 \pm 0.018$, yields a firm lower limit to 
the luminosity (mass) ratio of $\sim$100:1. Interestingly, our best estimate for 
the luminosity of the progenitor (based on the color of shell B) agrees 
remarkably well with the combined luminosity of the four shells.  This implies 
that most of the mass of the disrupted satellite actually ended up in the four 
shells that we detected.

What is the nature of the shell progenitor?  Adopting the results 
based on our analysis of shell B, the disrupted galaxy had an absolute 
$I$-band magnitude of $-18.7$.  This luminosity suggests that the progenitor 
was brighter than a typical dwarf galaxy.  Considering the unique location of 
NGC~4889 in the cluster core and the types of galaxies that currently 
surround it, it is reasonable to assume that the progenitor was 
an early-type system.  Moreover, numerical simulations suggest that the 
formation of regular shell structures normally requires the progenitor to 
have a relatively low stellar velocity dispersion.  Putting all of these clues 
together, we propose that the progenitor was most likely a sizable disk 
galaxy, presumably a low-mass S0 or early-type spiral.

\section{Summary} 

We have discovered a system of four stellar shell features in 
NGC~4889, the BCG in the Coma cluster.  The shells are well aligned with the 
major axis of the  host galaxy.  The unique environment of NGC~4889 in the 
central region of the massive Coma cluster, combined with the non-detection of 
any dust features, suggests that the shells are structures formed through a
non-dissipative merger event. We present a detailed 2-D structural 
decomposition of \hst/ACS images of the BCG and its complex surroundings, 
including a careful determination of the global and local sky background with
the help of large-field images from SDSS.  These technical steps are crucial for
securing an accurate determination of the photometric properties of the shells.  
We have successfully measured not only the luminosities but also the colors of 
the shells.  The shells are found to be slightly bluer than the host galaxy at 
the same location.

Under the assumption that the shells were created by a small satellite galaxy 
that fell into NGC~4889, and that the progenitor system obeyed the present-day 
observed color-magnitude relation of the Coma cluster, we use the measured 
colors of the shells to infer the luminosity (mass) of the disrupted galaxy.  
We estimate that the shell progenitor galaxy had an $I$-band absolute magnitude 
of $-18.7$, roughly 90 times less luminous than NGC~4889. The 
combined luminosity of the four shells is roughly consistent with the 
progenitor luminosity inferred from our analysis of the CMR, which suggests 
that most of the mass of the disrupted galaxy remained in the shell 
structures. 

The techniques outlined in this study provide a novel strategy to use the 
photometric substructure of early-type galaxies to constrain the mass ratio 
and other useful physical properties of past minor merger events.

\acknowledgements
This work was supported by the Carnegie Institution for Science (LCH) and the 
China Scholarship Council (MG, SH), and under the National Natural Science
Foundation of China under grant 11133001 and 11273015 (MG, SH). MG and SH thank 
Prof. Q.-S. Gu and the School of Astronomy and Space Science of Nanjing 
University for providing long-term support. Funding for the SDSS and SDSS-II has 
been provided by the Alfred P. Sloan Foundation, the Participating Institutions, 
the National Science Foundation, the U.S. Department of Energy, the National
Aeronautics and Space Administration, the Japanese Monbukagakusho, the Max
Planck Society, and the Higher Education Funding Council for England. The SDSS
Web site is {\tt http://www.sdss.org}. This research has made use of the 
NASA/IPAC Extragalactic Database (NED) which is operated by the Jet Propulsion
Laboratory, California Institute of Technology, under contract with the National
Aeronautics and Space Administration. We thank the anonymous 
referee for helpful comments that improved this paper.



\end{CJK*}

\end{document}